\documentclass[aps,pre,twocolumn,superscriptaddress,showpacs,floatfix,amsmath,amssymb]{revtex4-1}

\usepackage{color}
\usepackage{dcolumn}
\usepackage{amsmath}
\usepackage{amssymb}
\usepackage{mathtools}
\usepackage{graphicx}
\usepackage{latexsym}
\usepackage{amsfonts}
\usepackage{amssymb}
\usepackage{comment}
\usepackage[colorlinks=true,pdfstartview=FitV,linkcolor=black,citecolor=black,urlcolor=blue]{hyperref}

\newcommand{\be}{\begin{equation}}  
\newcommand{\ee}{\end{equation}}

\newcommand\varmp{\mathbin{\vcenter{\hbox{%
  \oalign{\hfil$\scriptstyle-$\hfil\cr
          \noalign{\kern-.3ex}
          $\scriptscriptstyle({+})$\cr}%
}}}}

\def\hc{\text{h.c.}}
\def\ud{{\mathrm{d}}}

\def\w{{\omega}}
\def\im{{\mathrm{i}}}
\def\ex{{\mathrm{e}}}
\def\a{\alpha}
\def\b{\beta}

\def\g{\gamma}
\def\eps{\epsilon}
\renewcommand\Im{\operatorname{Im}}
\renewcommand\Re{\operatorname{Re}}

\begin{document}

% Title stuff
\title{Comparing the generalized Kadanoff-Baym ansatz with the full Kadanoff-Baym equations for an excitonic insulator out of equilibrium}

\author{Riku Tuovinen}
\email[]{riku.tuovinen@utu.fi}
\affiliation{QTF Centre of Excellence, Turku Centre for Quantum Physics, Department of Physics and Astronomy, University of Turku, 20014 Turku, Finland}
\author{Denis Gole\ifmmode \check{z}\else \v{z}\fi{}}
\affiliation{Center for Computational Quantum Physics (CCQ), The Flatiron Institute, 162 Fifth avenue, New York NY 10010}
\author{Martin Eckstein}
\affiliation{Department of Physics, University of Erlangen-N{\"u}rnberg, 91058 Erlangen, Germany}
\author{Michael A. Sentef}
\affiliation{Max Planck Institute for the Structure and Dynamics of Matter, Luruper Chaussee 149, 22761 Hamburg, Germany}

%%%%%%%%%%%%%%%%%%%%%%%%%%%%%%%%%%%%%%%%%%%%%%%%%%%%%%%%%%%%%%%%%%%%%%%%%%%%%%%%
%%%%%%%%%%%%%%%%%%%%%%%%%%%%%%%%%%%%%%%%%%%%%%%%%%%%%%%%%%%%%%%%%%%%%%%%%%%%%%%%

% Abstract
\begin{abstract}
We investigate out-of-equilibrium dynamics in an excitonic insulator~(EI) with a finite momentum pairing perturbed by a laser-pulse excitation and a sudden coupling to fermionic baths. The transient dynamics of the excitonic order parameter is resolved using the full nonequilibrium Green's function approach and the generalized Kadanoff--Baym ansatz (GKBA) within the second-Born approximation. The comparison between the two approaches after a laser pulse excitation shows a good agreement in the weak and the intermediate photo-doping regime. In contrast, the laser-pulse dynamics resolved by the GKBA does not show a complete melting of the excitonic order after a strong excitation. Instead we observe persistent oscillations of the excitonic order parameter with a predominant frequency given by the renormalized equilibrium bandgap. This anomalous behavior can be overcome within the GKBA formalism by coupling to an external bath, which leads to a transition of the EI system towards the normal state. We analyze the long-time evolution of the system and distinguish decay timescales related to dephasing and thermalization.
\end{abstract}

\maketitle

%%%%%%%%%%%%%%%%%%%%%%%%%%%%%%%%%%%%%%%%%%%%%%%%%%%%%%%%%%%%%%%%%%%%%%%%%%%%%%%%
%%%%%%%%%%%%%%%%%%%%%%%%%%%%%%%%%%%%%%%%%%%%%%%%%%%%%%%%%%%%%%%%%%%%%%%%%%%%%%%%

\section{Introduction}\label{sec:intro}

Quantum dynamics out of equilibrium can be used to disentangle interesting mechanisms of materials' properties, such as origin of ordered states and their subsequent control. Recent experimental progress in pump-probe-spectroscopical approaches to excitonic insulator~\cite{Mor2017}, charge-density wave~\cite{Zong2018}, and superconducting phases~\cite{Fausti2011,Mitrano2016} has prompted extensive research interest in both simulating~\cite{Werner2012, Eckstein2013, Golez2016, Okamoto2016, Sentef2017, Murakami2017a, Murakami2017b, Babadi2017, Claassen2017, Kennes2017, Mazza2017a, Nava2017, Herrmann2017, Fabrizio2018, Li2018, Werner2018, Perfetto2019PRM, Perfetto2020, Perfetto2020PRB} and measuring~\cite{Kaiser2014, Hu2014, Denny2015, Mitrano2016, Werdehausen2018, Dendzik2020, Baldini2020} ultrafast quantum correlation effects far from equilibrium.

Simulating these processes can be challenging since an accurate but computationally feasible theoretical description is required for simultaneously dealing with strong external fields, many-particle interactions, and transient effects. The nonequilibrium Green's function (NEGF) approach can address all these challenges~\cite{Danielewicz1984,svlbook,Balzer2013book}: It is not limited to weak driving or linear response only, the many-particle correlations can be systematically included by construction of self-energy diagrams, and the real-time Green's function gives access to time-dependent observables such as densities, currents, total energies, and spectral functions. The drawback is in the computational effort for solving the dynamical equations of motion for the Green's function, which scale with the number of timesteps cubed. A simplification to this issue was proposed already over 30 years ago in Ref.~\cite{Lipavsky1986} by reducing the two-time-propagation of the Green's function to the time-propagation of a time-local density matrix via the generalized Kadanoff-Baym ansatz (GKBA), thereby reducing the computational scaling to the number of timesteps squared. While this approach was acknowledged and used already in the 1990s~\cite{Haug1992,Tso1992,Bonitz1996,Jahnke1997,Kwong1998}, its recent revival~\cite{Galperin2008, Ness2011, Hermanns2012, Balzer2012, Balzer2013, Hermanns2014, BarLev2014, Latini2014, Schluenzen2017, Hopjan2018, Perfetto2018cheers, Covito2018a, Covito2018b, Karlsson2018, Tuovinen2019pssb, Hopjan2019, Schueler2019, Murakami2020} has made it possible to combine the NEGF approach with \textit{ab initio} descriptions of realistic atomic, molecular, and condensed matter systems~\cite{Perfetto2015, Bostrom2018, Perfetto2018, Perfetto2019, Schueler2020}. Recent development has further allowed for an equivalent but more efficient representation of the GKBA time evolution with only a linear scaling in the number of timesteps~\cite{Schluenzen2020,Joost2020,Karlsson2020}.

In this work, we consider ultrafast many-particle correlations in an excitonic-insulator system acting as a prototypical ordered-phase material~\cite{Golez2016,Mor2017,Murakami2017a,Mazza2020}. Out-of-equilibrium dynamics in such systems with a symmetry-broken ground state has been shown to be extremely sensitive to all the intricacies in the electronic and lattice structure~\cite{Golez2016,Murakami2017a,Tanabe2018,Tuovinen2019pssb,Baldini2020}. We drive the system out of equilibrium in two ways: (1) by an external laser pulse, and (2) by coupling to fermionic baths. We compare the resolved dynamics for the NEGF between the full Kadanoff-Baym equations (KBE) and the computationally less expensive GKBA. We find that while the laser-pulse excitation introduces rich transient dynamics with predominant oscillations given by a renormalized bandgap, the GKBA description, in contrast to KBE, does not damp to a stationary solution. This can be attributed to narrow spectral features of the GKBA, the character of the approximation for the propagators, and correlation-induced damping in the KBE solution~\cite{vonFriesen2010}. Coupling to fermionic baths instead opens up a natural decay channel for the GKBA description as well, and we observe clear damping and even a transition from the excitonic to the normal state. We further characterize the nature of this phase transition by identifying separate decay timescales.

The paper is organized as follows. In Section~\ref{sec:model} we introduce the model system, and we outline the main equations of the NEGF and GKBA approach. The out-of-equilibrium dynamics due to external laser pulses and coupling to fermionic baths are analyzed in Section~\ref{sec:results}. In Section~\ref{sec:concl} we summarize our conclusions and discuss future prospects.

\section{Model and method}\label{sec:model}

\subsection{Model for the excitonic insulator}

\begin{figure}[t]
\centering
\includegraphics[width=0.45\textwidth]{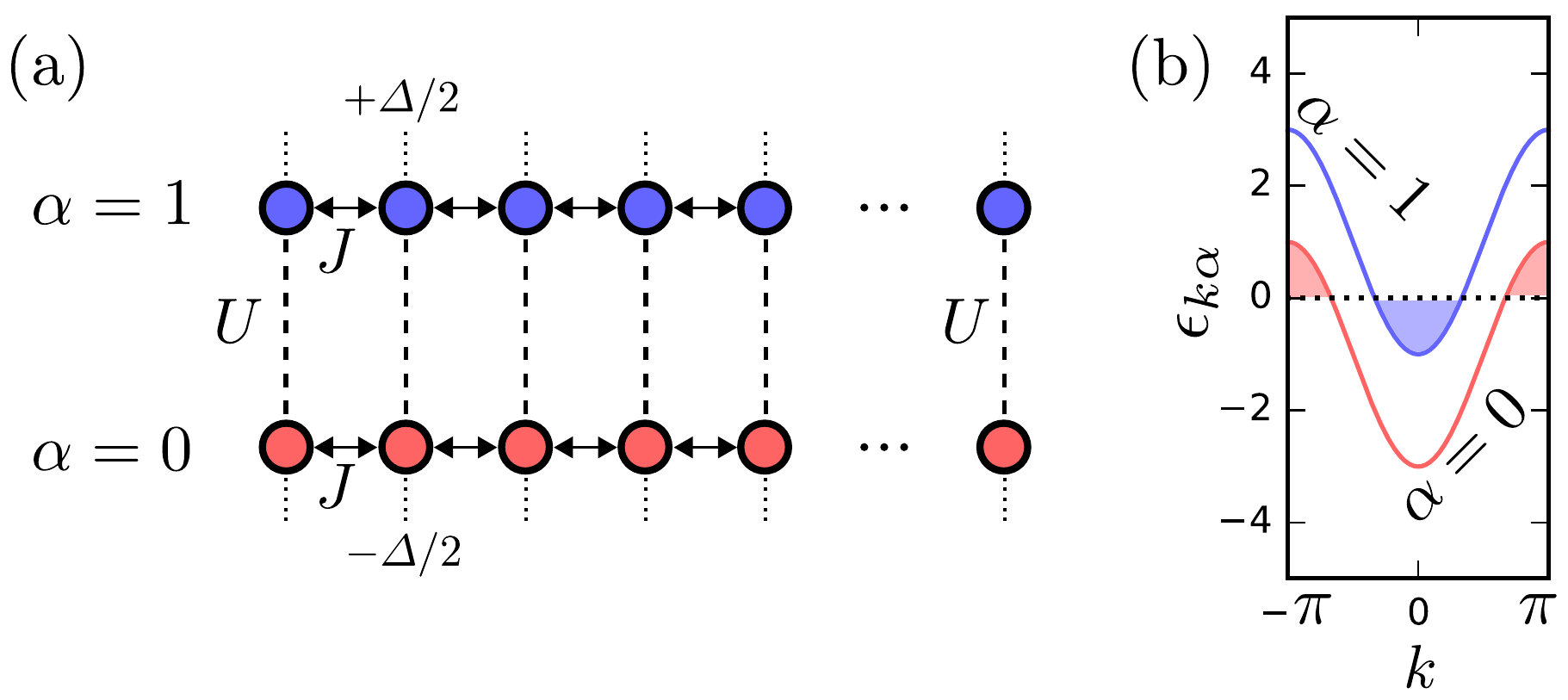}
\caption{Model system schematic. (a) One-dimensional chains with nearest-neighbor hopping $J$, on-site energy $\pm \varDelta/2$, and interband Hubbard interaction $U$. (b) Noninteracting bandstructure of the two separate chains with $J=-1$ and $\varDelta=2$. Filling of the bands is set by the chemical potential at zero (dashed line) describing a hole pocket at the band edge for the lower band and an electron pocket at the band center for the upper band (shaded areas).}
\label{fig:schematic}
\end{figure}

We model the excitonic insulator (EI) by a two-band system of spinless fermions~\cite{Golez2016,Tuovinen2019pssb}
\be\label{eq:h0}
\hat{H}_0 = \sum_{k\a} (\eps_{k\a} + \varDelta_\a) \hat{d}_{k\a}^\dagger \hat{d}_{k\a} ,
\ee
where $\hat{d}_{k\a}^{(\dagger)}$ are the annihilation (creation) operators for electrons with momentum $k$ in band $\a\in\{0,1\}$ labeling the two bands, and $\varDelta_\a$ is the associated crystal field leading to the bandgap $\varDelta \equiv |\varDelta_0 - \varDelta_1|$. In practice, we consider a real-space structure of two one-dimensional chains with periodic boundary condition, see Fig.~\ref{fig:schematic}(a). Each of these two real-space structures result in each of the two bands as seen in Fig.~\ref{fig:schematic}(b). The creation and annihilation operators in momentum and real space are related by $\hat{d}_{k\alpha}^{(\dagger)} = (1/\sqrt{N_\a})\sum_{m} \exp[\varmp\im k m]\hat{c}_{m\a}^{(\dagger)}$, where $m\a$ labels the real-space lattice site $m$ of the chain $\a\in\{0,1\}$. For one-dimensional chains with nearest-neighbor hopping $J_\a$ the energy band dispersion is $\eps_{k\a} = 2 J_\a \cos k$. In this picture, the crystal field $\varDelta_\a$ can be readily identified as the local on-site energy for the lattice points. In the real-space picture the Hamiltonian in Eq.~\eqref{eq:h0} then reads
\be\label{eq:h0lattice}
\hat{H}_0 = \sum_{mn\a} h^0_{m\a, n\a}\hat{c}_{m\a}^\dagger \hat{c}_{n\a},
\ee
where the matrix elements are chosen such that for nearest neighbors in each chain $h^0_{m\a, n\a}=J$ and for on-site $h^0_{m\a, m\a}=\varDelta_\a$ with $\varDelta_{0(1)}=\varmp \varDelta/2$. For all calculations in the present work, we set $J=-1$ and calculate energies in units of $|J|$ and times in units of $|J|^{-1}$.

The electron-electron interaction is taken as an interband Hubbard interaction~\cite{Golez2016,Tuovinen2019pssb}
\be\label{eq:v}
\hat{H}_{\text{int}} = U \sum_m\hat{c}_{m0}^\dagger\hat{c}_{m0}^\dagger\hat{c}_{m1}\hat{c}_{m1} ,
\ee
introducing a local density-density interaction of strength $U$ for the electrons between the two bands. The electron-electron interaction is the origin of excitonic pairing between an electron pocket at $k=0$ and a hole pocket at $k=\pi$, see Fig.~\ref{fig:schematic}(b). The excitonic insulator phase is determined by a finite order parameter $\langle \hat{d}_{(k+\pi)0}^\dagger \hat{d}_{k1}\rangle \neq 0$ which spontaneously breaks the conservation of charge in each of the bands and the spatial symmetry. We discuss the practical evaluation of the order parameter in Sec.~\ref{sec:obs}. The pairing introduces a finite hybridization between the bands and opens a gap.

An external laser pulse driving the above system out of equilibrium is modeled by a direct transition between the two bands~\cite{Golez2016}
\be\label{eq:hext}
\hat{H}_{\text{ext}}(t) = A(t)\sum_{k} ( \hat{d}_{k1}^\dagger \hat{d}_{k0} + \hc ) ,
\ee
where we set the pulse shape as a gaussian: $A(t)=A \sin[\w(t-t_c)]\ex^{-4.6(t-t_c)^2/t_c^2}$ of amplitude $A$, frequency $\w$, and centering $t_c = 2\pi n_p/\w$ with $n_p$ being the number of optical cycles. Using the transformation introduced below Eq.~\eqref{eq:h0} we rewrite also Eq.~\eqref{eq:hext} in real space. Since the laser-pulse term couples the two bands at equal $k$-points, using the property $\sum_{k\in[-\pi,\pi)} \ex^{\im k (m-n)}=\delta_{mn}$ we obtain a straightforward replacement
\be\label{eq:hextlattice}
\hat{H}_{\text{ext}}(t) = A(t)\sum_m ( \hat{c}_{m1}^\dagger \hat{c}_{m0} + \hc ).
\ee

The total Hamiltonian for the above setup combining the kinetic, interaction, and external terms then reads
\be\label{eq:htot}
\hat{H} = \hat{H}_0 + \hat{H}_{\text{int}} + \hat{H}_{\text{ext}} .
\ee
From now on, we use matrix representations of these objects in terms of the one-particle states in the real-space basis $\{|m\a\rangle\}$: $\langle m\a | \hat{H}_0 + \hat{H}_{\text{ext}}(t) | n\b \rangle = h_{m\a, n\b}(t)$ and $\langle m\a | \hat{H}_{\text{int}} | n\b \rangle = v_{m\a, n\b}(t)$. While the interaction term itself is instantaneous, in Eq.~\eqref{eq:v}, we allow the strength of it to be time-dependent to describe adiabatic switching, which we will discuss in Section~\ref{sec:results}.

\subsection{Time propagation of the nonequilibrium Green's functions}

We employ the nonequilibrium Green's function (NEGF) method where the Kadanoff--Baym equations are propagated in time~\cite{Baym1961,kbbook,Keldysh1964,Danielewicz1984,Danielewicz1984b,
Kohler1999,Semkat1999,Kwong2000,Myohanen2008,Stan2009b,Myohanen2009,
vonFriesen2009,Balzer2010,svlbook,Balzer2013book,Schluenzen2016,Schluenzen2019}.
The key quantity in the formalism is the one-particle Green's function, which we write in the one-particle basis of our model system,
\be
G_{m\a,n\b}(z,z') = -\im \left\langle \mathcal{T}_\g \left[ \hat{c}_{m\a,\mathrm{H}}(z) \hat{c}_{n\b,\mathrm{H}}^\dagger(z') \right] \right\rangle ,
\ee
where $z,z'$ are time coordinates on the Keldysh contour $\g$ with the contour-time-ordering operator $\mathcal{T}_\g$. The contour $\g$ has a forward branch, $z=t_- \in [0,\infty)$, and a backward branch, $z=t_+ \in (\infty,0]$, on the real-time axis, and also a vertical branch, $z=-\im\tau\in[0,-\im\beta]$ on the imaginary axis, see e.g.~\cite{svlbook}. Here we set, without loss of generality, the contour starting point at zero on the real-time axis, $z\equiv t=0$. The creation and annihilation operators are represented in the Heisenberg picture, and the ensemble average, denoted by $\langle\cdots\rangle$, is taken as a trace over
the density matrix. The Green's function matrix $G(z,z')$ is the solution to the integro-differential equation of motion (in matrix form)
\be\label{eq:kbe}
[ \im\partial_z - h(z) ]G(z,z') = \delta(z,z') + \int_\g \ud\bar{z} \ \varSigma[G](z,\bar{z})G(\bar{z},z'),
\ee
where $h(z)$ is the one-particle Hamiltonian for the system, $\delta(z,z')$ is a delta function on the Keldysh contour, and $\varSigma[G]$ is the self-energy kernel containing all the information about many-particle and embedding effects. The integration is performed over the Keldysh contour through the Langreth rules~\cite{Langreth1972,Langreth1976}. Depending on the contour-time arguments, $(z,z')$, the double-time functions appearing in Eq.~\eqref{eq:kbe} can be represented in components: lesser ($<$), greater ($>$), retarded (R), advanced (A), left ($\lceil$), right ($\rceil$), and Matsubara (M)~\cite{svlbook}. The self-energy kernel $\varSigma[G]$ can be obtained from an underlying $\varPhi$-functional, $\varSigma[G] = \delta \varPhi[G] / \delta G$, to guarantee the satisfaction of various macroscopic conservation laws~\cite{Baym1962}, provided that the equations of motion are solved self-consistently~\cite{vonBarth1996,Holm1998,Dahlen2006,Stan2009a}.

The Green's function provides a direct access to system observables such as densities and currents of the out-of-equilibrium system. In particular, we are interested in the time-dependent one-particle reduced density matrix (TD1RDM) given by the time-diagonal of the lesser Green's function, $\rho(t)\equiv -\im G^<(t,t)$. At the equal-time limit on the real-time axis, $z=t_-$, $z' \to t_+$, we obtain from Eq.~\eqref{eq:kbe} and its adjoint~\cite{Latini2014,Hopjan2018,Perfetto2018cheers,Tuovinen2019pssb}
\be\label{eq:rhoeom}
\frac{\ud}{\ud t} \rho(t) + \im [h(t)+\varSigma_{\text{HF}}(t),\rho(t)] = -[I(t)+\hc],
\ee
where we separated the self-energy, $\varSigma(t,t')\equiv \varSigma_{\text{HF}}(t)\delta(t,t') + \varSigma_{\text{corr}}(t,t')$, in time-local Hartree--Fock (HF) and time-non-local correlation (corr) contributions, and we also introduced the collision integral in terms of the correlation part~\cite{Latini2014,Hopjan2018,Perfetto2018cheers,Tuovinen2019pssb}
\be\label{eq:collint}
I(t) = \int_0^t \ud \bar{t} [ \varSigma_{\text{corr}}^>(t,\bar{t})G^<(\bar{t},t) - \varSigma_{\text{corr}}^<(t,\bar{t})G^>(\bar{t},t) ] .
\ee 
We use the one-particle basis of our model system to write the self-energy at the HF level~\cite{Balzer2010,Tuovinen2019pssb}
\begin{align}\label{eq:sigmahf}
(\varSigma_{\text{HF}})_{m\a, n\b}(t) & = \delta_{mn}\delta_{\a\b} \sum_{p\zeta} v_{m\a, p\zeta}(t) \rho_{p\zeta, p\zeta}(t) \nonumber \\
& - v_{m\a, n\beta}(t) \rho_{n\beta, m\a}(t) ,
\end{align}
and the correlation self-energy at the second-order Born (2B) level~\cite{Balzer2010,Tuovinen2019pssb}
\begin{align}\label{eq:sigma2b}
(\varSigma_{\text{corr}})^\lessgtr_{m\a, n\b}(t,t') & = \sum_{p\zeta q\eta} v_{m\a, p\zeta}(t) v_{n\b, q\eta}(t') G^\gtrless_{q\eta, p\zeta}(t',t) \nonumber \\
& \times \left[ G^\lessgtr_{m\a, n\b}(t,t')G^\lessgtr_{p\zeta, q\eta}(t,t') \right.\nonumber \\
& \left. - \ G^\lessgtr_{m\a, q\eta}(t,t')G^\lessgtr_{p\zeta, n\b}(t,t') \right] .
\end{align}
We note that since our model describes spinless fermions, the spin-degeneracy factor~\cite{Balzer2010,Balzer2013book}, typically written for the direct terms [first terms on the right-side of Eqs.~\eqref{eq:sigmahf} and~\eqref{eq:sigma2b}], is here simply $1$.

The combination of the equation of motion in Eq.~\eqref{eq:kbe} and the expressions of the self-energies in Eqs.~\eqref{eq:sigmahf} and~\eqref{eq:sigma2b} represents a closed set of equations for the full solution based on KBE. We solve these equations using the numerical library NESSi~\cite{Schueler2019Nessi}. In particular, we solve the problem in momentum space and use a suitable MPI parallelization over momentum points, see Ref.~\cite{Golez2016} for details. In the full KBE solution, the collision integral in Eq.~\eqref{eq:collint} also includes the initial-correlation part on the imaginary branch of the Keldysh contour $\sim \int_0^\beta \ud \tau \varSigma_{\text{corr}}^\rceil (t,\tau) G^\lceil(\tau,t)$~\cite{Schueler2019Nessi}. From now on, we refer to this approach as 2B@KBE.

An alternative approach to close the equation of motion for $\rho$ in Eq.~\eqref{eq:rhoeom} is to employ the GKBA approximation~\cite{Lipavsky1986,Hermanns2012}
\be\label{eq:gkba}
G^\lessgtr(t,t') \approx \mp G^{\text{R}}(t,t')\rho^\lessgtr(t') \pm \rho^\lessgtr(t)G^{\text{A}}(t,t'),
\ee
where we denoted $\rho^< \equiv \rho$ and $\rho^> \equiv 1-\rho$, and we represent the retarded/advanced propagators at the HF level~\cite{Hermanns2012,Hermanns2014}
\be\label{eq:propagator}
G^{\text{R}/\text{A}}(t,t') = \mp \im \theta[\pm (t-t')] \mathcal{T} \ex^{-\im \int_{t'}^t \ud \bar{t}[ h(\bar{t})+\varSigma_{\text{HF}}(\bar{t})]} 
\ee
with $\mathcal{T}$ being the chronological time-ordering operator. We then use Eq.~\eqref{eq:gkba} in Eqs.~\eqref{eq:sigma2b} and~\eqref{eq:collint}, and then solve for the TD1RDM in Eq.~\eqref{eq:rhoeom} by using a time-stepping algorithm~\cite{Stan2009a,Tuovinen2019pssb}. While the inclusion of initial correlations has been shown to be possible also within GKBA~\cite{Semkat2003,Karlsson2018,Hopjan2019,Bonitz2019}, here we adiabatically switch on the many-particle interactions and only include the collision integral in the form of Eq.~\eqref{eq:collint}. For efficient computation, we additionally use a recurrence relation for constructing Eq.~\eqref{eq:propagator} due to its group property~\cite{Balzer2013book,Tuovinen2019pssb}, and we employ optimized matrix (tensor) operations for the construction of the 2B self-energy~\cite{Tuovinen2019-2B}. From now on, we refer to this approach as 2B@GKBA.

\subsection{Inclusion of fermionic baths}

So far we have considered isolated systems being exposed to external drives locally within the system. Now we add a contribution from a bath environment, e.g., a particle reservoir or a biased electrode, described by~\cite{svlbook,Tuovinen2013,Tuovinen2014}
\be\label{eq:hbath}
\hat{H}_{\text{bath}}(z) = \sum_{k\lambda} \varepsilon_{k\lambda}(z) \hat{b}_{k\lambda}^\dagger \hat{b}_{k\lambda} ,
\ee
where $k\lambda$ labels the $k$-th basis function in the $\lambda$-th bath. The bath energy dispersion depends on the Keldysh contour time $z$~\cite{Ridley2015,Ridley2017}
\be\label{eq:voltage}
\varepsilon_{k\lambda}(z) = \begin{cases} \varepsilon_{k\lambda} - \mu & \text{when} \ z\equiv t<0 \\ \varepsilon_{k\lambda} + V_\lambda(t) & \text{when} \ z\equiv t\geq 0 ,\end{cases}
\ee
where $\mu$ is the equilibrium chemical potential and $V_\lambda(t)$ is a generic excitation, such as a bias voltage, taking place at $z \equiv t=0$. The bath is coupled to the EI system by the coupling Hamiltonian~\cite{svlbook,Tuovinen2013,Tuovinen2014}
\be\label{eq:hcoupling}
\hat{H}_{\text{coupling}}(z) = \sum_{m\a k\lambda} [J_{m\a, k\lambda}(z) \hat{c}_{m\a}^\dagger \hat{b}_{k\lambda} + \hc ] ,
\ee
where $J_{m\a, k\lambda}$ are the coupling matrix elements between the EI system and the bath, which in general also depend on the Keldysh contour time $z$. In this work, we consider the ``partitioned approach''~\cite{Stefanucci2004,Ridley2018} where the systems are brought in contact at $z \equiv t=0$. These contributions in Eqs.~\eqref{eq:hbath} and~\eqref{eq:hcoupling} are then added to the total Hamiltonian in Eq.~\eqref{eq:htot}.

We consider electronic interactions only within the EI system. Hence, for a noninteracting bath environment the relevant Green's functions are given by~\cite{svlbook,Tuovinen2013,Tuovinen2014,Ridley2015}
\begin{align}
g_{k\lambda}^{\text{R}/\text{A}}(t,t') & = \mp \im \theta [ \pm (t-t')] \ex^{-\im \int_{t'}^t \ud \bar{t} [\varepsilon_{k\lambda}+V_\lambda(\bar{t})]} \label{eq:grbath} \\
g_{k\lambda}^{\lessgtr}(t,t') & = \pm \im f[\pm(\varepsilon_{k\lambda}-\mu)]\ex^{-\im\int_{t'}^t \ud \bar{t} [\varepsilon_{k\lambda}+V_\lambda(\bar{t})] }
\end{align}
where $f(x)=1/(\ex^{\b x}+1)$ is the Fermi function at inverse temperature $\b$, and we used $f(-x) = 1-f(x)$.

We may then readily write the retarded/advanced bath self-energy, which is completely specified by the bath and coupling Hamiltonians~\cite{svlbook,Tuovinen2013,Tuovinen2014,Ridley2015},
\begin{align}\label{eq:sigmabath}
& \varSigma_{\text{bath},\lambda}^{\text{R}/\text{A}}(t,t') \nonumber \\
& = \ex^{-\im \psi_\lambda(t,t')}\int \frac{\ud \w}{2\pi} \ex^{\im \w (t-t')} \left[\varLambda_\lambda(\w) \mp \im \varGamma_\lambda(\w)/2\right] ,
\end{align}
where we introduced $\psi_\lambda(t,t')\equiv \int_{t'}^t \ud \bar{t} V_\lambda(\bar{t})$ and~\cite{Ridley2015}
\begin{align}
(\varLambda_{\lambda})_{m\a, n\b}(\w) & = \sum_k J_{m\a, k\lambda} \mathcal{P}\left(\frac{1}{\w-\varepsilon_{k\lambda}}\right) J_{k\lambda, n\b} , \label{eq:lambda}\\
(\varGamma_{\lambda})_{m\a, n\b}(\w) & = 2\pi\sum_k J_{m\a, k\lambda} \delta(\w-\varepsilon_{k\lambda})J_{k\lambda, n\b} , \label{eq:gamma}
\end{align}
and we used the Cauchy relation for the relative-time Fourier transform of Eq.~\eqref{eq:grbath}, $1/(\w-\varepsilon_{k\lambda}\pm \im \eta) = \mathcal{P}(1/(\w-\varepsilon_{k\lambda})) \mp \im\pi\delta(\w-\varepsilon_{k\lambda})$, with $\eta$ being a positive infinitesimal and $\mathcal{P}$ denoting the principal value~\cite{svlbook}. It is important to notice that the bath self-energy is represented in the basis of the EI system because it describes the effect of ``embedding'' the EI system into the bath environment.
We now assume the frequency content of the bath self-energy is much broader than the energy scales in the EI system, known as the wide-band approximation (WBA). This approximation is justified here as we are concentrating on very low-energy excitations within the EI system at which the bath density of states is practically featureless~\cite{Zhu2005,Verzijl2013,Covito2018c,Ridley2019}.
%In practice this could be achieved using substrates or electrodes with large bandwidths, e.g., bulk gold.
In the WBA, the level-width matrix becomes independent of frequency, $\varGamma_\lambda(\w) \approx \varGamma_\lambda$, which means it becomes time-local. Then, also the real part of the self-energy in Eq.~\eqref{eq:lambda} vanishes due to Kramers--Kronig relations. Thus, the retarded/advanced bath self-energy is obtained by further summing over the bath index $\lambda$~\cite{svlbook,Tuovinen2013,Tuovinen2014,Ridley2015}
\begin{align}\label{eq:sigma-bath-retarded-advanced}
\varSigma_{\text{bath}}^{\text{R}/\text{A}}(t,t') & = \sum_\lambda \varSigma_{\text{bath},\lambda}^{\text{R}/\text{A}}(t,t') = \mp \frac{\im}{2} \sum_\lambda \varGamma_\lambda \delta(t-t') \nonumber \\
& = \mp \frac{\im}{2} \varGamma \delta(t-t') .
\end{align}
Similarly, we obtain for the lesser/greater bath self-energy~\cite{Croy2009,Tuovinen2013,Ridley2015}
\begin{align}\label{eq:sigma-bath}
& \varSigma_{\text{bath}}^{\lessgtr}(t,t') \nonumber \\
& = \pm \im \sum_\lambda \varGamma_\lambda \ex^{-\im\psi_\lambda(t,t')} \int \frac{\ud \w}{2\pi} f[\pm(\w-\mu)]\ex^{-\im\w(t-t')} .
\end{align}
Due to the WBA, the frequency integral in Eq.~\eqref{eq:sigma-bath} as such is not convergent but we use a cutoff frequency, $\w_c$, based on the physical band edge of the bath given by the bath energy dispersion: $\varGamma_\lambda \to \varGamma_\lambda(\w) = \theta(\w_{c}-|\w|)\varGamma_\lambda$.

Since the retarded/advanced bath self-energy was obtained as a time-local contribution in Eq.~\eqref{eq:sigma-bath-retarded-advanced}, it can directly be included in the HF propagators in Eq.~\eqref{eq:propagator}~\cite{Latini2014,Perfetto2018cheers}
\be\label{eq:propagator-g}
G^{\text{R}/\text{A}}(t,t') = \mp \im \theta[\pm (t-t')] \mathcal{T} \ex^{-\im \int_{t'}^t \ud \bar{t} [h(\bar{t})+\varSigma_{\text{HF}}(\bar{t})\mp \im \varGamma/2]}.
\ee
The lesser/greater component of the bath self-energy in Eq.~\eqref{eq:sigma-bath}, in contrast, appears in an additional collision integral~\cite{Latini2014,Perfetto2018cheers}
\begin{align}
& I_{\text{bath}}(t) \nonumber \\
& = \int_0^t \ud \bar{t} [\varSigma_{\text{bath}}^>(t,\bar{t})G^<(\bar{t},t) - \varSigma_{\text{bath}}^<(t,\bar{t})G^>(\bar{t},t)] ,\label{eq:bath-collint}
\end{align}
whose contribution is added to Eq.~\eqref{eq:collint}. Also, the GKBA of Eq.~\eqref{eq:gkba} is used for the lesser/greater Green's functions in Eq.~\eqref{eq:bath-collint}.

\subsection{Accessing physical observables}\label{sec:obs}

The TD1RDM, $\rho(t)$, as a solution to Eq.~\eqref{eq:rhoeom} naturally contains the information about the single-particle density on its diagonal, but also time-dependent expectation values of any single-particle operator $\hat{O}$ may be extracted using it by~\cite{Joost2019}
\be
\langle \hat{O} \rangle(t) = -\im \sum_{mn} O_{m,n} \rho_{n,m}(t).
\ee

In our model system, we consider excitonic pairing between an electron pocket of the upper band (around $k=0$) and a hole pocket of the lower band (around $k= \pm \pi$), see Fig.~\ref{fig:schematic}(b). In practice, this means that in the EI phase $\langle \hat{d}_{(k+\pi)0}^\dagger \hat{d}_{k1}\rangle \neq 0$. Therefore, we average this object over the reduced Brillouin zone (RBZ), $\sum_{k\in[-\pi/2,\pi/2)} \equiv \sum_k'$, and define this as the excitonic order parameter~\cite{Golez2016,Tuovinen2019pssb}
\begin{align}\label{eq:orderparameter}
\phi (t) & \equiv \frac{1}{N_k} \sum_k {}' \langle \hat{d}_{(k+\pi)0}^\dagger \hat{d}_{k1}\rangle \nonumber \\
& = \frac{2}{N} \sum_{m,n=1}^{N/2} (-1)^m f_{mn} \rho_{m , (n+N/2)} (t)
\end{align}
where $N_k$ is the number of $k$ points in the RBZ, $N$ is the total number of real-space lattice points, and we introduced
\be
f_{mn} \equiv \frac{1}{N_k}\sum_k {}' \ex^{\im k (m-n)} \stackrel{N_k\to \infty}{\longrightarrow} \frac{\sin [\frac{\pi}{2} (m-n)]}{\frac{\pi}{2} (m-n)},
\ee
where the limiting case applies for infinite lattice sites. In practice, we evaluate the RBZ sum numerically, but in most cases already $N=20$ corresponds to the sinc function fairly reasonably. On the second line of Eq.~\eqref{eq:orderparameter} we used the transformation of the field operators between momentum and real-space, which also results in the alternating sign, $(-1)^m = (\ex^{\im \pi})^m$. Momentum-averaged band populations could be obtained similarly.

The total energy in the system can be divided in three contributions: (1) single-particle (or kinetic) energy $E_{\text{single}}(t) = \Re\text{Tr}[h(t) \rho(t)]$, where $h$ includes the single-particle Hamiltonian and the external field; (2) HF energy $E_{\text{HF}}(t) = \frac{1}{2}\Re\text{Tr}[\varSigma_{\text{HF}}(t) \rho(t)]$ corresponding to the time-local part; and (3) correlation energy $E_{\text{corr}}(t) = -\frac{1}{2}\Im \text{Tr}[I(t)]$ being the remaining part of the collision integral after removing the HF part~\cite{Balzer2013book}. While the effect of exchanging energy between the EI system and the external bath could be included in this description, we perform the energy considerations only for the isolated system. The total energy then reads
\be\label{eq:etot}
E_{\text{tot}}(t) = E_{\text{single}}(t) + E_{\text{HF}}(t) + E_{\text{corr}}(t) .
\ee
We can further calculate energy absorption during some time interval by the difference
\be\label{eq:eabs}
E_{\text{abs}} = E_{\text{tot}}(t_{\text{final}}) - E_{\text{tot}}(t_{\text{initial}}) ,
\ee
where $t_{\text{final}}$ is, e.g., the total propagation time, and $t_{\text{initial}}$ the time when some external fields are being switched on. Alternatively, this could also be evaluated from a Hellmann-Feynman formula $E_{\text{abs}}=\int_{t_{\text{initial}}}^{t_{\text{final}}} \ud t' \partial_{t'} A(t') 2 \mathrm{Re} \sum_k \langle \hat{d}^{\dagger}_{k1} \hat{d}_{k0} \rangle$, since the field depends explicitly on time but the expectation value only implicitly.

\begin{figure}[t]
\centering
\includegraphics[width=0.5\textwidth]{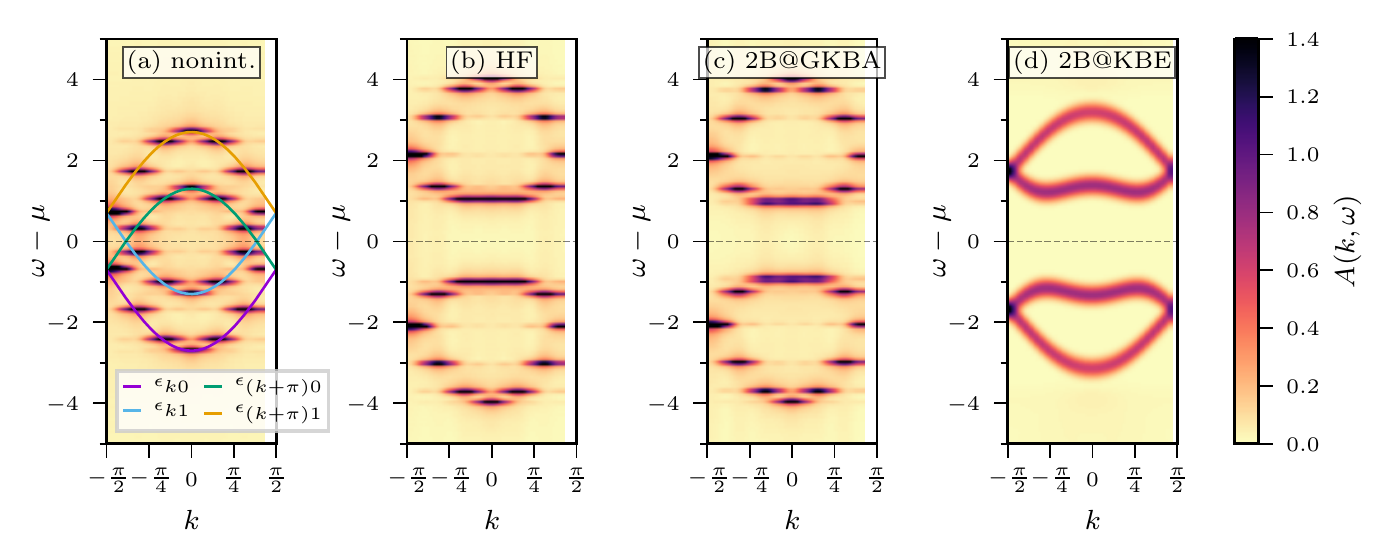}
\caption{Energy- and momentum-resolved equilibrium spectral function for the (a) noninteracting system, and for the interacting systems described at the (b) Hartree-Fock, (c) second-Born level with the GKBA, and (d) second-Born level with the full KBE. The EI system parameters are $\varDelta=1.4$, $U=\{0.0,3.5\}$. The reduced Brillouin zone for the momentum axis is shown using back-folding. The noninteracting energy-band structure is superimposed onto the noninteracting spectral function in panel (a) with solid lines.}
\label{fig:spectral-eq}
\end{figure}

The nonequilibrium spectral function is defined as~\cite{Latini2014}
\be
A(t,t') \equiv \im \left[ G^{\text{R}}(t,t') - G^{\text{A}}(t,t') \right] ,
\ee
which is a matrix in the one-particle states of our model system. It is important to note that the GKBA in Eq.~\eqref{eq:gkba} satisfies the exact condition $G^{\text{R}}-G^{\text{A}} = G^> - G^<$. We then calculate a spatio-temporal Fourier transformation of the nonequilibrium spectral function with respect to the real-space lattice coordinates and the relative-time coordinate $\tau \equiv t-t'$~\cite{Joost2019}
\begin{align}\label{eq:spectral}
A(k,\w) & = \frac{\im}{N}\sum_{m n} \ex^{\im k (m - n)} \int \ud \tau \ex^{\im\w\tau} \nonumber \\
& \times [ G_{m, n}^>(T+\frac{\tau}{2},T-\frac{\tau}{2}) - G_{m, n}^<(T+\frac{\tau}{2},T-\frac{\tau}{2}) ] ,
\end{align}
where $N$ is the total number of lattice points and $T \equiv (t+t')/2$ is the center-of-time coordinate. In practice, we evaluate it by setting $T$ to half the total propagation time, when the relative-time coordinate $\tau$ spans the maximal range diagonally in the two-time plane. Eq.~\eqref{eq:spectral} can be used to obtain the full energy dispersion or the bandstructure.  It is worth mentioning that while the spectral features obtained this way within the GKBA are limited by the choice of propagators at the HF level [cf. Eq.~\eqref{eq:propagator}], the lesser and greater Green's functions still include effects at the 2B@GKBA level.

Using Eq.~\eqref{eq:spectral} we show the equilibrium spectral functions of the EI system (with system parameters $\varDelta=1.4$, $U=\{0.0,3.5\}$) in Fig.~\ref{fig:spectral-eq} using both the GKBA and the full KBE approach. In the GKBA data we have used $N=24$ as the total number of lattice points, hence the energy bands consist of discrete peaks, in contrast to the $k$-resolved KBE data in Fig.~\ref{fig:spectral-eq}(d). In the limit of infinite number of lattice sites, these would produce the continuum energy-band structure of the EI system. In equilibrium we see the gap opening due to the excitonic condensate, see Fig.~\ref{fig:spectral-eq}(b). The energy axis is adjusted with the equilibrium chemical potential to take the Hartree shift into account. We also see that the 2B@GKBA equilibrium spectral function, obtained via the adiabatic switching procedure to be discussed in the next Section, is very similar to the HF one: The density of states is modified slightly but the overall structure remains. Importantly, the 2B@KBE spectral features are more broadened compared to 2B@GKBA.

\section{Results}\label{sec:results}

\subsection{Driving with a laser pulse}

\begin{figure}[t]
\centering
%\showthe\columnwidth
\includegraphics[width=0.5\textwidth]{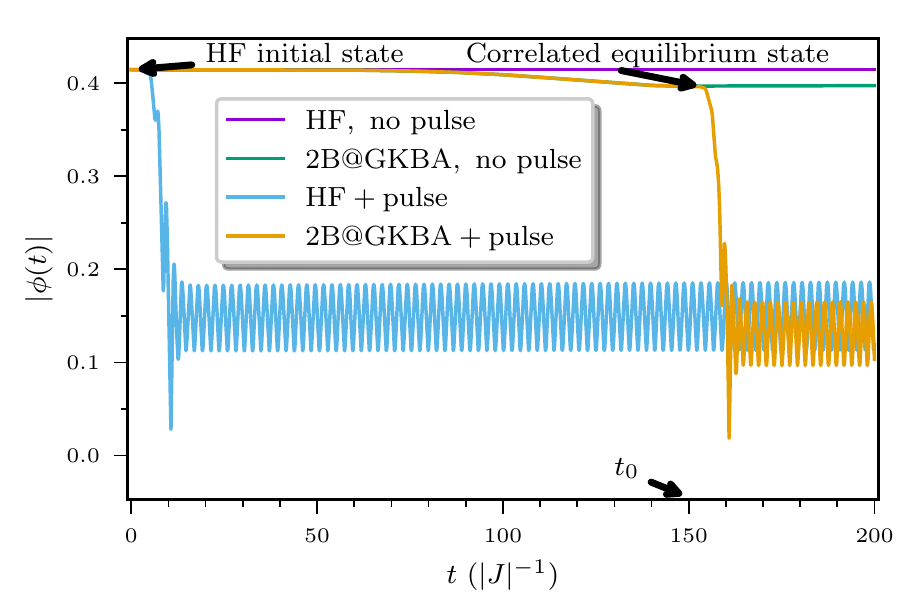}
\caption{Time evolution of the absolute value of the order parameter $|\phi(t)|$ with and without the excitation for the HF and 2B@GKBA propagation scheme. For the 2B@GKBA case the adiabatic preparation of the correlated equilibrium state is exemplified and followed by the application of the laser pulse.}
\label{fig:op-pulse}
\end{figure}

For all calculations, we consider our system to be in the EI phase by setting $\varDelta=1.4$ and $U=3.5$~\cite{Tuovinen2019pssb}. In Fig.~\ref{fig:op-pulse} we exemplify the generic procedure for the time-dependent simulations. For the description of interactions at the HF level, the initial equilibrium state can be obtained with a separate time-independent calculation~\cite{Tuovinen2019pssb}, and consequently the out-of-equilibrium behavior can readily be analyzed starting from $t=0$. Here, we are mainly interested in the description of interactions at the 2B@GKBA level, going beyond the mean-field description. For this analysis, we first need to prepare the correlated equilibrium state. This can be obtained by an initial time evolution ($t<t_0$) without external fields but adiabatically switching on the many-particle interactions in the 2B@GKBA self-energies~\cite{Tuovinen2019pssb}. After this, the out-of-equilibrium behavior, due to a laser excitation for example, can be studied ($t\geq t_0$). We note in passing that the preparation step may consume a considerable amount of computational time~\cite{Tuovinen2019pssb}, and it would be highly attractive to apply some sort of a restart protocol, e.g. of Refs.~\cite{Semkat2003,Karlsson2018,Hopjan2019,Bonitz2019}, for a separate calculation starting at $t=t_0$ including the initially correlated state. However, we have experienced in numerous tests (not shown) for this procedure to result in non-stationary behavior. We suspect the EI system considered here, possessing a symmetry-broken ground state with nonzero coherences on the off-diagonals of the density matrix~\cite{Tuovinen2019pssb}, may not provide an applicable equilibrium state, at least in the context of Ref.~\cite{Karlsson2018}.

\begin{figure}[t]
\centering
\includegraphics[width=0.5\textwidth]{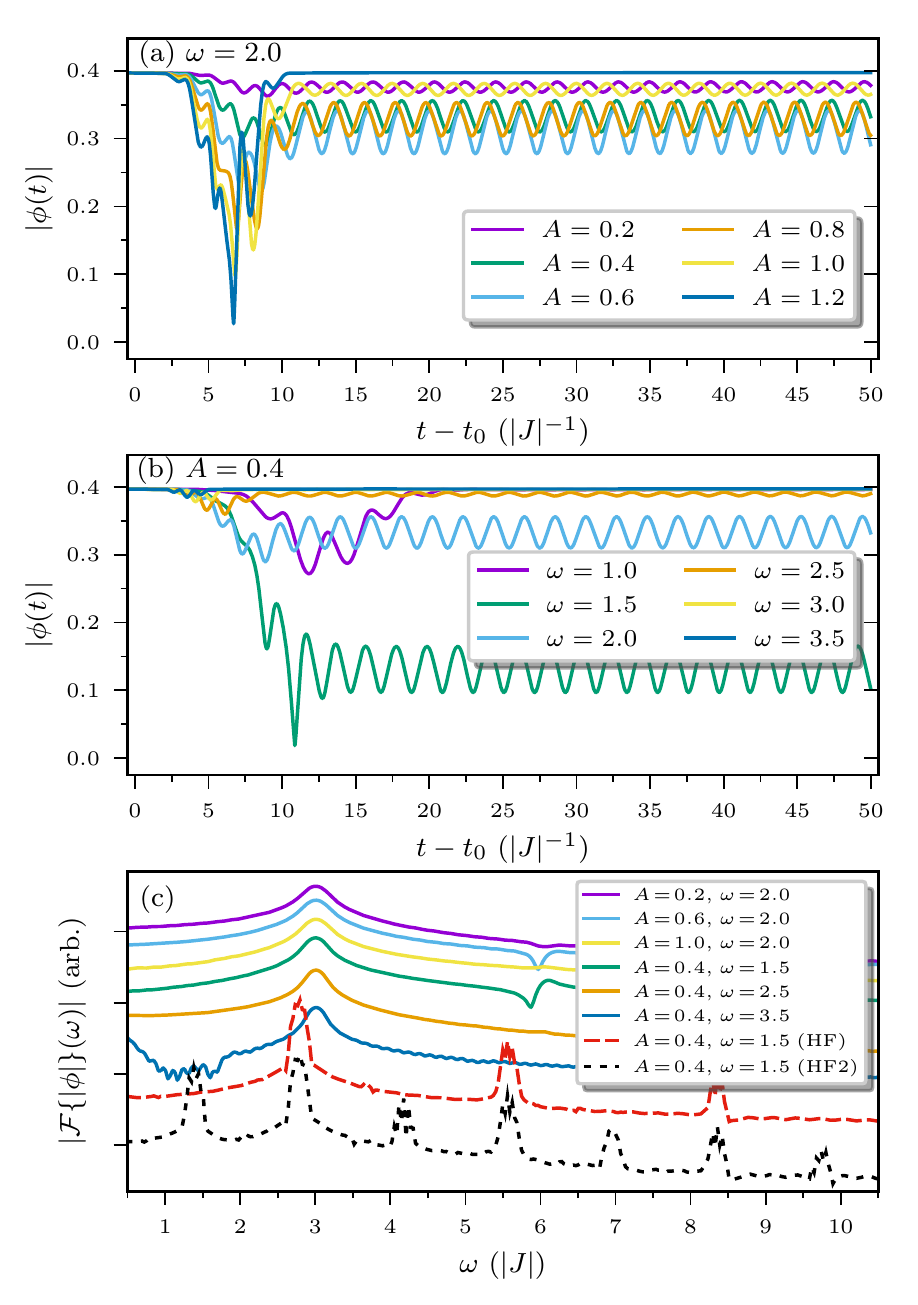}
\caption{Time evolution of the absolute value of the excitonic order parameter $|\phi(t)|$ after applying a laser pulse with (a) fixed frequency and varying amplitude, and (b) fixed amplitude and varying frequency. (c) Fourier spectra of selected time-dependent data from panels (a) and (b) for the 2B@GKBA propagation~(full lines). The Fourier spectrum for the HF solutions are marked with the dashed line and the HF2 represents the spectrum for lattice model with next-nearest neighbor hopping, see text for details. The curves are shifted vertically for clarity.}
\label{fig:panels-pulse}
\end{figure}

Let us first look at a concrete example of the time evolution at the HF or 2B@GKBA level. We fix the number of optical cycles in the laser pulse for all simulations $n_p=2$, cf. Eq.~\eqref{eq:hext}. In Fig.~\ref{fig:op-pulse}, we see that for the HF evolution the absolute value of the order parameter $|\phi(t)|$ remains constant without the applied field and it is substantially reduced and oscillating after the photo-excitation~($A=0.4$, $\w=1.5$). On the level of 2B@GKBA, the adiabatic switching procedure keeps the system in the EI phase, which is stationary without the applied field. This condition might change for different values of $U$ and $\varDelta$~\cite{Tuovinen2019pssb}. When we apply the the laser excitation the out-of-equilibrium dynamics is roughly similar in HF and 2B@GKBA: In 2B@GKBA the oscillation frequency is slightly increased compared to HF (see also Fig.~\ref{fig:panels-pulse}(c) and the consequent discussion later on). Next, we will focus on the 2B@GKBA case and thoroughly analyze how the EI system's response depends on the laser excitation.

\begin{figure}[t]
\centering
\includegraphics[width=0.5\textwidth]{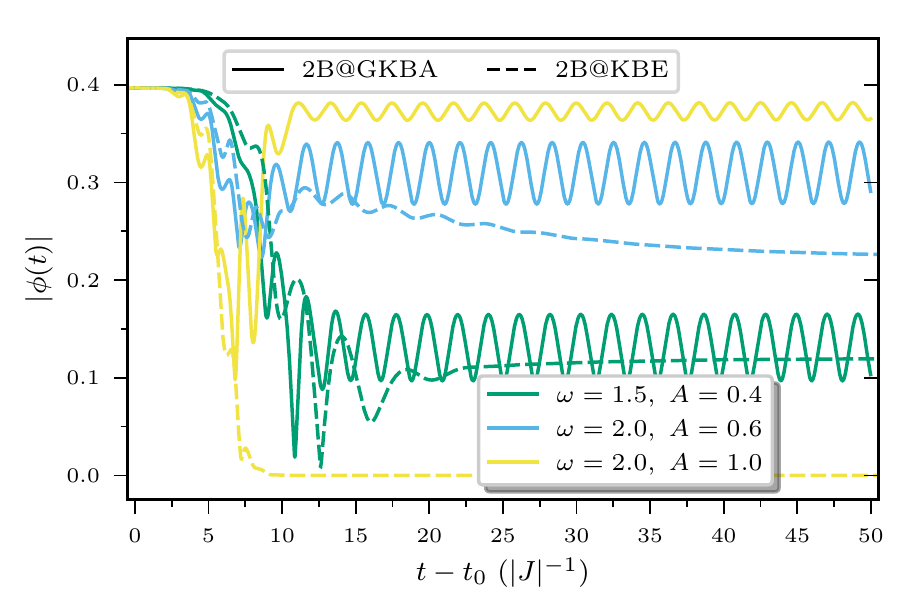}
\caption{Comparison of the absolute value of the excitonic order parameter $|\phi(t)|$ evolution within the 2B@GKBA~(solid lines) and the 2B@KBE~(dashed lines) propagation scheme.}
\label{fig:comparison-pulse}
\end{figure}

Stronger driving amplitude in the laser pulse expectedly makes the initial transient response stronger. This can be seen in Fig.~\ref{fig:panels-pulse}(a) for $t-t_0 \lesssim 6 \ |J|^{-1}$ where the excitonic order parameter decreases rapidly from its equilibrium value. This, however, does not mean the excitonic condensate would melt completely. Instead, the order parameter remains at an oscillatory but nonzero steady-state value after the laser pulse.
The frequency of these steady-state oscillations is independent of the driving amplitude as can be seen from the Fourier spectra in Fig.~\ref{fig:panels-pulse}(c) and corresponds to the amplitude mode excitations. The Fourier spectra are calculated using Blackman-window filtering~\cite{Blackman1959}. As we increase the excitation strength, namely $A \geq 1$, the order parameter after the photo-excitation is, somewhat counterintuitively, negligibly reduced. We will address this point more thoroughly later on.

The system expectedly responds more strongly to the resonant driving. This is seen in Fig.~\ref{fig:panels-pulse}(b) where we find the system to be most in resonance with the driving frequency $\w=1.5$. However, while the 2B@GKBA solution properly describes the resonance condition, it still retains its oscillatory character because of the lack of damping in the HF propagators. The oscillations of the excitonic order parameter after the laser excitation are independent of the laser frequency as can be seen from the Fourier spectra in Fig.~\ref{fig:panels-pulse}(c). We also show the Fourier spectra of the HF data (cf.~Fig.~\ref{fig:op-pulse}). As we saw already in Fig.~\ref{fig:op-pulse} the oscillation frequency in 2B@GKBA is slightly increased compared to HF, from $2.8$ to approximately $3$. These values can be attributed to the equilibrium system parameter for the noninteracting bandgap $\varDelta=1.4$ as we see even harmonics with frequencies $2n\varDelta$ (with $n$ a positive integer) in the HF spectrum. The oscillation can therefore be associated with the crystal field; even though the bandstructure gets modified due to the electron-electron interaction, cf.~Fig.~\ref{fig:spectral-eq}, the transient signatures include the remnants of the crystal field. We can verify this finding by breaking the symmetry of our lattice model by introducing a next-nearest-neighbor hopping $J'=J/2$ (HF2 in Fig.~\ref{fig:panels-pulse}(c)), in which case also the odd harmonics appear with frequencies $(2n+1)\varDelta$. In the 2B@GKBA data, the higher order harmonics are more suppressed while the basic resonant frequency, related to a renormalized equilibrium bandgap, remains clearly visible in all cases independent of the laser amplitude or frequency.

We compare the 2B@GKBA solution to that of the full 2B@KBE in Fig.~\ref{fig:comparison-pulse}. In the weak excitation regime $A \lesssim 1$, the excitonic order parameter is nonzero in the long time limit and its value roughly agrees between the 2B@GKBA and 2B@KBE results. However, the 2B@KBE solution shows a considerably stronger damping than the one of 2B@GKBA. This is due to the quasi-particle corrections beyond HF, in contrast to the form in Eq.~\eqref{eq:propagator}, and the consequent correlation-induced damping~\cite{vonFriesen2010}. For instance, if the driving frequency is slightly off-resonant, namely $\w=2.0$, the narrow spectral window of 2B@GKBA does not capture as much of the weight as the more broadened 2B@KBE which damps towards a slightly different steady-state value. In case of the resonant driving $\w=1.5$, the reduction of the order parameter is in an excellent agreement between the 2B@GKBA and 2B@KBE results. On the other hand, the dynamics is qualitatively different for strong excitation strengths $A \geq 1.0$. While in the 2B@GKBA the order is negligibly reduced, it is completely melted for the 2B@KBE propagation scheme and the EI system undergoes a transition to the normal state consistent with the $GW$ level description reported in Ref.~\cite{Golez2016}.

\begin{figure}[t]
\centering
\includegraphics[width=0.5\textwidth]{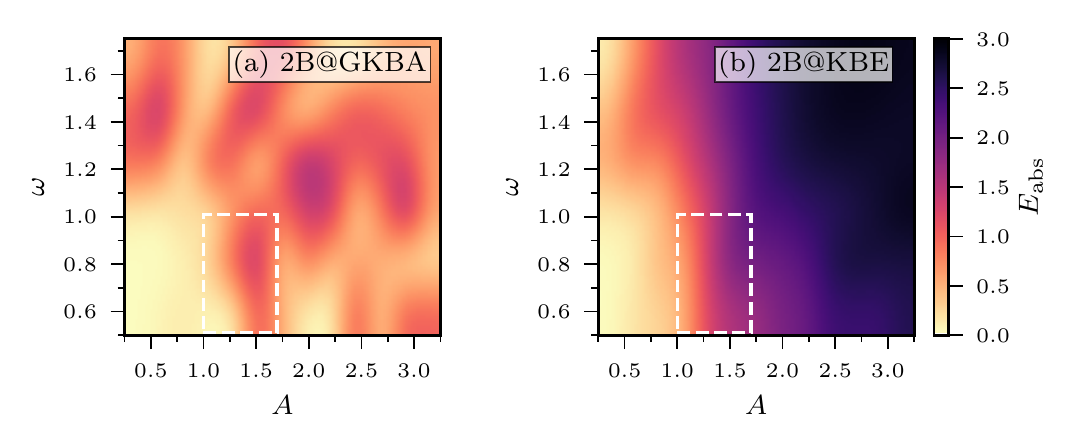}
\caption{Energy-absorption spectrum (color map) in terms of the laser pulse amplitude (horizontal axis) and frequency (vertical axis) for (a) the 2B@GKBA and for (b) the full 2B@KBE solution. See text for the discussion of the dashed region.}
\label{fig:eabs}
\end{figure}

The dependence on the driving amplitude presents the main difference between the 2B@GKBA and the 2B@KBE solution. Within the 2B@GKBA the steady-state value of the order parameter may depend nontrivially on the driving amplitude. For instance, for pulse frequency $\w=2$ the order parameter is maximally reduced around $A=0.6$ in Fig.~\ref{fig:panels-pulse}(a). Higher amplitude pulses seem not to break the electron-hole pairs, keeping the excitonic order parameter roughly at its equilibrium value. This means that how the laser pulse get absorbed to the EI system depends strongly on the width of the spectral features, which are more narrow in 2B@GKBA than in 2B@KBE, see Fig.~\ref{fig:spectral-eq}. We analyze this behavior more in detail in Fig.~\ref{fig:eabs}, where we show the energy absorption calculated using Eq.~\eqref{eq:eabs} as a function of the driving amplitude and frequency for both the 2B@GKBA and the full 2B@KBE solution. We have checked (not shown) that possible finite-size effects in 2B@GKBA are negligible as larger number of lattice sites in the EI model leads to qualitatively similar data. For both cases, we observe that for smaller driving amplitudes ($A \lesssim 1$) the energy absorption is expectedly maximal around the resonant frequency $\w=1.5$ related to the renormalized equilibrium bandgap, cf.~Fig.~\ref{fig:panels-pulse}. However, for 2B@GKBA, if we follow a line at fixed frequency, e.g., at $\w=1.5$, we see that the energy absorption oscillates with the driving amplitude. This is not the case for the full 2B@KBE solution, where higher-amplitude pulses straightforwardly lead to larger absorption. For the 2B@KBE solution the moderately large electron-electron interaction $U=3.5$ gives already considerable broadening, resulting in energy absorption and consequently melting of the excitonic condensate at any amplitude $A \gtrsim 1.5$ (cf.~Fig.~\ref{fig:panels-pulse}). On the other hand, we may conclude that the 2B@GKBA description is reasonable at weak fields close to resonance, but this picture breaks down at stronger fields off-resonance due to nonlinear absorption and higher order scattering mechanisms.

\begin{figure}[t]
\centering
\includegraphics[width=0.5\textwidth]{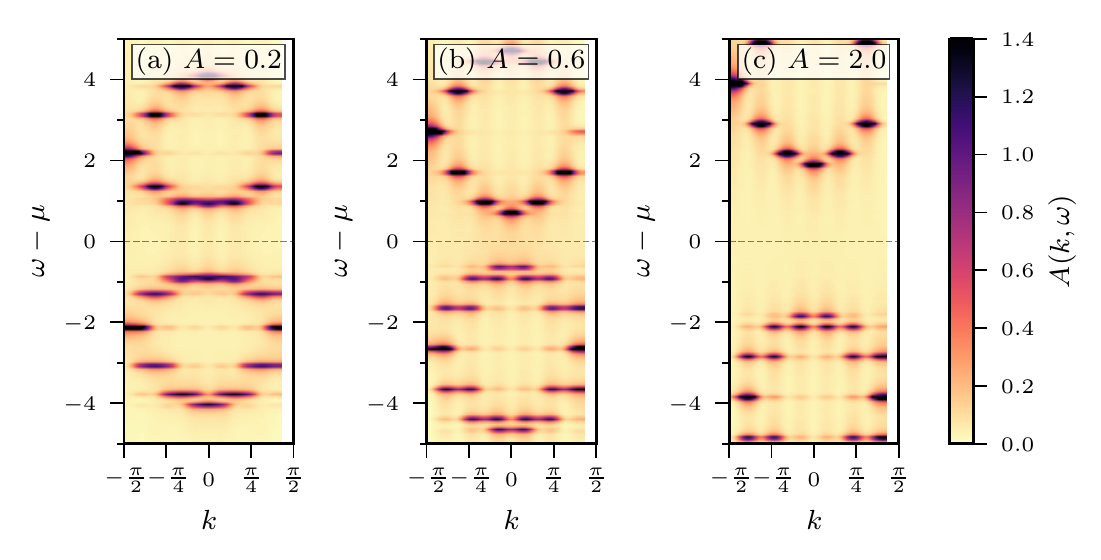}
\caption{Energy- and momentum-resolved equilibrium spectral function at the 2B@GKBA level with a constant dipolar transition term of strength (a) $A=0.2$, (b) $A=0.6$, and (c) $A=2.0$.}
\label{fig:spectral-const}
\end{figure}

An interesting observation in the analysis of the absorbed energy is a softening of the absorption edge with an increased excitation strength, see the dashed regions in Fig.~\ref{fig:eabs}. For $A\lesssim 1.5$ this onset of nonlinear absorption also seems consistent between 2B@GKBA and 2B@KBE. We can understand this phenomenon by analyzing a static problem with a constant dipolar matrix element.
Because the form of the excitation in Eq.~\eqref{eq:hext} introduces a direct dipolar transition matrix element, $\langle\hat{d}_{k1}^\dagger\hat{d}_{k0}\rangle$, it pushes the lowest and highest bands away from each other which, in turn, moves the backfolded bands in the middle closer to each other, cf.~Fig.~\ref{fig:spectral-eq}(a). The electron-electron interaction, on the other hand, introduces a further coupling between the bands in the middle, $\langle\hat{d}_{(k+\pi)0}^\dagger\hat{d}_{k1}\rangle$, leading to a competition between the excitonic order and the dipolar matrix element. We can verify this behavior by looking at the energy- and momentum-resolved spectral function in Fig.~\ref{fig:spectral-const}. In this calculation, we consider the equilibrium system supplemented with a constant dipolar transition $A$ as in Eq.~\eqref{eq:hext}, which then shows how the bandstructure would be affected by this form of an excitation, in general. While these equilibrium spectral functions do not exactly correspond to the laser-pulse situation, it provides us with some insight on the underlying mechanism. We see the gap closing around $A=0.6$, which is in this case the critical point where the equilibrium system transforms from the excitonic to the normal state. Higher transition amplitudes introduce simply a rigid shift of the bands away from each other when the electron-hole interaction is no longer binding them together. It would also be feasible to calculate the nonequilibrium spectral function due to the short laser-pulse excitation. However, due to the competing mechanisms and in contrast to Fig.~\ref{fig:spectral-const}, it would show a very rich and complex spectrum of multiple photon-assisted side bands, and as clear interpretation as in Fig.~\ref{fig:spectral-const} would be challenging.

\subsection{Coupling to fermionic baths}

We now consider each lattice site of the two chains in our EI system to be coupled to two different baths with equal coupling strength $J_{m\a,k\lambda}$ in Eq.~\eqref{eq:hcoupling}. As the level width or tunneling rate $\varGamma$ in Eq.~\eqref{eq:gamma} depends not only on the coupling strength but also on the bath energy dispersion, we investigate the role of bath coupling by directly varying the strength of $\varGamma$. The bath filling is modified by a bias $V_\lambda(t)$ in Eq.~\eqref{eq:voltage} which we set to a constant value $\varmp V$ for the bath connected to the $\a=0$ ($\a=1$) chain of the EI system. For the bath environment we additionally fix $\b=100$ in Eq.~\eqref{eq:sigma-bath}. This effectively resembles a zero-temperature limit at which the adiabatic switching procedure is consistent.

\begin{figure}[t]
\centering
\includegraphics[width=0.5\textwidth]{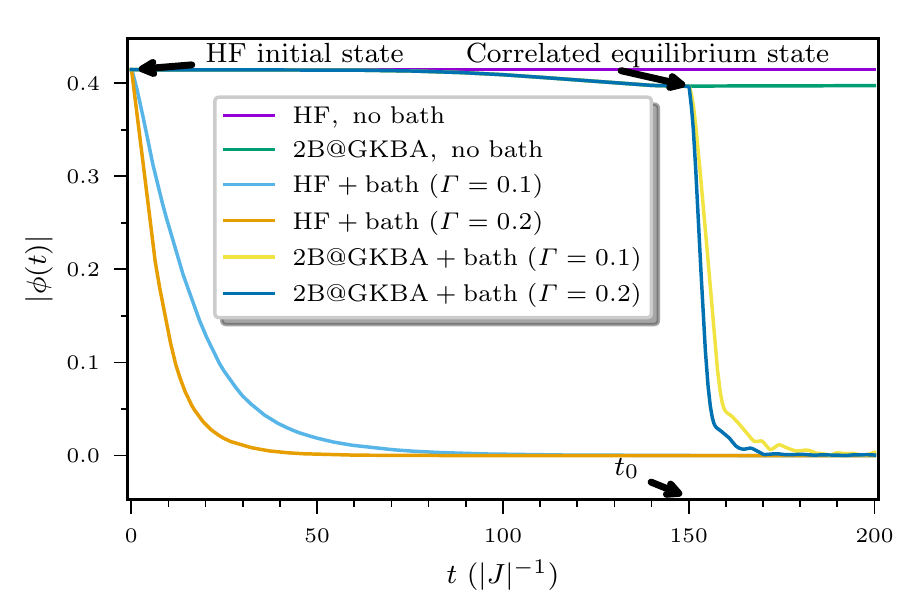}
\caption{Time evolution of the absolute value of the order parameter $|\phi(t)|$ with and without the application of the fermionic bath for the HF and 2B@GKBA propagation scheme. For the 2B@GKBA case, the adiabatic preparation of the correlated equilibrium state is exemplified and followed by the application of the bath coupling.}
\label{fig:op-bath}
\end{figure}

The procedure for analyzing the dynamics induced by the bath coupling is similar to the laser pulse excitation in the previous subsection. We first prepare the correlated equilibrium state by the adiabatic switching procedure~\cite{Tuovinen2019pssb} and then suddenly bring the system in contact with the baths. The excitonic order parameter responds to this external perturbation as seen in Fig.~\ref{fig:op-bath}. Also in this case, for a description of the electronic correlations at the HF level only, the bath coupling could be introduced without the preparation step, and the corresponding dynamics shows only a straightforward decay process depending on the coupling strength between the EI system and the baths, see Fig.~\ref{fig:op-bath}. This decay behaviour is drastically modified when the electronic correlations are described at the 2B@GKBA level. Next, we will analyze this in detail by looking at the dynamics after the bath coupling at $t_0=150$ when varying (1) the bias, (2) the bath coupling duration, and (3) the bath coupling strength.

\begin{figure}[t]
\centering
\includegraphics[width=0.5\textwidth]{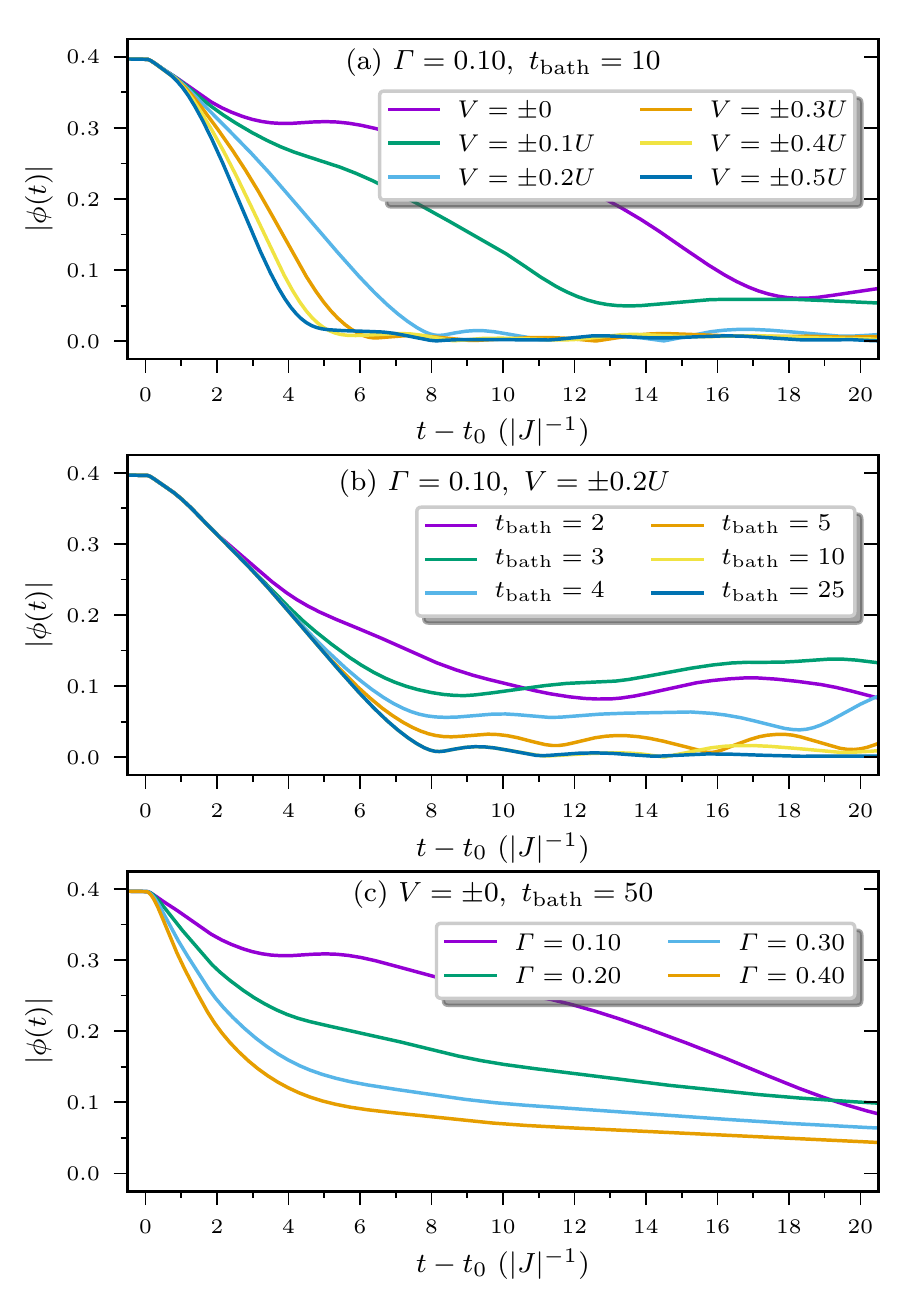}
\caption{Time evolution of the excitonic order parameter after the bath coupling with (a) fixed tunneling rate $\varGamma=0.1$, fixed coupling duration $t_{\mathrm{bath}}=10$, and varying bias $V$; (b) fixed tunneling rate $\varGamma=0.1$, fixed bias $V=\pm 0.2U$, and varying coupling duration $t_{\mathrm{bath}}$; (c) fixed bias $V=\pm 0$, fixed coupling duration $t_{\mathrm{bath}}=50$, and varying tunneling rate $\varGamma$.}
\label{fig:panels-bath}
\end{figure}

The bias changes the overall decay timescale of the excitonic condensate. In Fig.~\ref{fig:panels-bath}(a) we fix the bath coupling strength $\varGamma=0.1$ and the bath coupling duration $t_{\text{bath}}=10$, and we show the excitonic order parameter dynamics when the bias is increased from $V=\pm 0$ to $V=\pm 0.5U$. The final state can have nonzero excitonic order if the energy injected by the bias is not large enough to break the electron-hole pairs completely. However, even the bath coupling itself without bias lowers the order parameter compared to the equilibrium value. The initial transient at $t-t_0 < 2 \ |J|^{-1}$ is completely specified by the bath coupling strength, and the consequent decay dynamics depends on the bias.

The bath coupling duration does not change the overall decay timescale of the excitonic condensate. In Fig.~\ref{fig:panels-bath}(b) we fix the bath coupling strength $\varGamma=0.1$ and the bias $V=\pm 0.2U$, and we expose the EI system to the baths for varying durations. The initial transient on all the curves collapses onto one decay process described by the bath coupling strength and the bias, see also Fig.~\ref{fig:panels-bath}(a). The final state can also in this case have nonzero excitonic order if the bath exposure duration is short enough, but a transition from the EI state to a normal state is introduced for longer exposure durations.

Increasing the bath coupling strength, while keeping the bias and exposure duration fixed, makes the system undergo a faster decay process towards the normal state, see Fig.~\ref{fig:panels-bath}(c). This is understandable since stronger bath coupling directly influences the exponential decay timescale in Eq.~\eqref{eq:propagator-g}. However, for weaker couplings the initial transient shows competing mechanisms for breaking and recombining electron-hole pairs. Interestingly, we also observe multiple exponential decay timescales which will analyze in detail next.

\begin{figure}[t]
\centering
\includegraphics[width=0.5\textwidth]{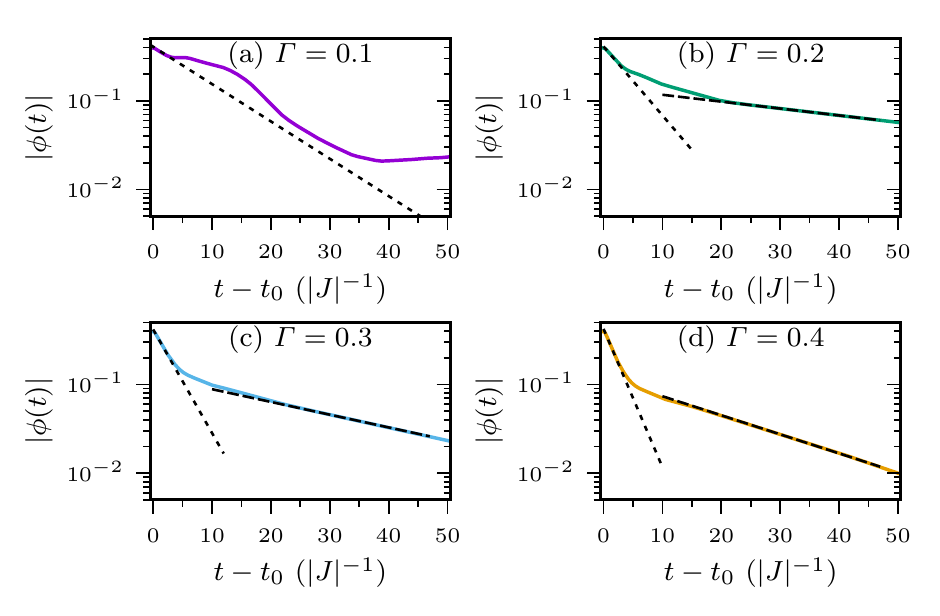}
\caption{Exponential fits for the decay timescales of the excitonic order parameter shown in Fig.~\ref{fig:panels-bath}(c). The short dashed lines are specified completely by $\sim \ex^{-\varGamma t}$ while the long-dashed lines are obtained by fitting to the flat part (except in panel (a)).}
\label{fig:timescales}
\end{figure}

We show the decay timescales of Fig.~\ref{fig:panels-bath}(c) separately in Fig.~\ref{fig:timescales} in logarithmic scale, and we see clearly that the initial transient in all the cases is also here completely specified by the bath coupling strength. We thereby refer to this mechanism as \textit{dephasing}~\cite{tsuji2013,babadi2015,Golez2016}. A second exponential decay process can be seen when the bath coupling is strong enough to melt the excitonic condensate completely related to \textit{thermalization}~\cite{tsuji2013,babadi2015,Golez2016}. In this case, the bias was fixed to $V=\pm 0$, and the thermalization appears slower than dephasing. However, as we have seen in Fig.~\ref{fig:panels-bath}(a) the bias will affect the overall decay timescale, and increasing the bias can also make the thermalization faster than dephasing. We will look closer into this effect next.

\begin{figure}[t]
\centering
\includegraphics[width=0.5\textwidth]{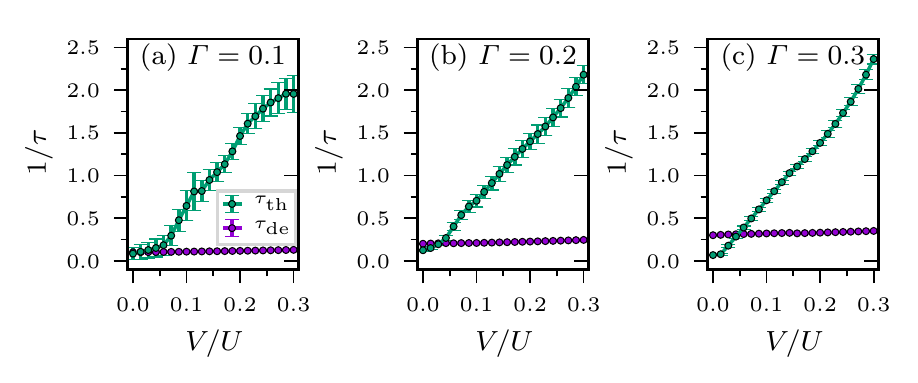}
\caption{Decay timescale exponents ($\sim \ex^{-t/\tau}$) as a function of the bias $V$ and varying tunneling rate (a) $\varGamma=0.1$, (b) $\varGamma=0.2$, and (c) $\varGamma=0.3$. The statistical error bars are given by the numerical fitting procedure; for all $\tau_{\mathrm{de}}$ datapoints the error is smaller than the marker size.}
\label{fig:expvsfilling}
\end{figure}

In Fig.~\ref{fig:expvsfilling} we show the numerically extracted decay exponents from a wide selection of simulated decay processes with varying bias and coupling. We see that the dephasing timescale, $\tau_{\mathrm{de}}$, remains roughly constant (given directly by the bath coupling strength $1/\tau_{\mathrm{de}} \approx \varGamma$) while the thermalization timescale, $\tau_{\mathrm{th}}$, is affected by the bias. The trend here is consistent with Fig.~\ref{fig:panels-bath}(a) where we observed that higher bias results in faster decay. This is also similar to Ref.~\cite{Golez2016} where $\tau_{\mathrm{th}}$ reportedly grows with the excitation strength. 

\begin{figure}[t]
\centering
\includegraphics[width=0.5\textwidth]{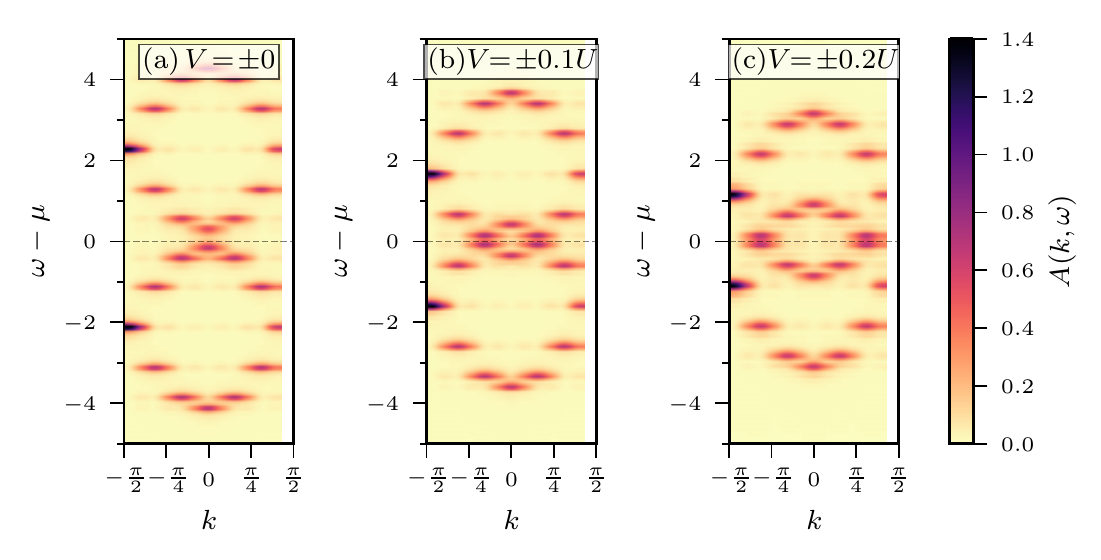}
\caption{Energy- and momentum-resolved nonequilibrium spectral function after coupling the baths with a fixed tunneling rate $\varGamma=0.1$ and varying bias (a) $V=\pm 0$, (b) $V=\pm 0.1U$, and (c) $V=\pm 0.2U$.}
\label{fig:spectral-bath}
\end{figure}

We can gain some more insight into these decay timescales by looking at the energy- and momentum-resolved nonequilibrium spectral function in Fig.~\ref{fig:spectral-bath}. Compared to the equilibrium spectral function in Fig.~\ref{fig:spectral-eq} the bath coupling expectedly modifies the spectral features drastically. In Fig.~\ref{fig:spectral-bath}(a) we see that already with zero bias the coupled system's gap starts closing. For larger bias [Fig.~\ref{fig:spectral-bath}(b) and~\ref{fig:spectral-bath}(c)] the system evidently transforms towards the normal state, cf.~Fig.~\ref{fig:spectral-eq}(a).
It is also interesting to note that compared to the excitation in Fig.~\ref{fig:spectral-const}, the spectral properties in the case of bath coupling, Fig.~\ref{fig:spectral-bath}, 
behave more straightforwardly as there seem to be no competing effects. This picture also translates into the clean decay dynamics of the excitonic condensate seen in Fig.~\ref{fig:panels-bath} and the disentangled decay timescales seen in Figs.~\ref{fig:timescales} and~\ref{fig:expvsfilling}.

\section{Conclusion}\label{sec:concl}

We have considered the out-of-equilibrium dynamics in a prototypical ordered-phase material, namely the excitonic insulator. We have studied out-of-equilibrium conditions due to a laser-pulse excitation and coupling the EI system to a fermionic bath. The calculations based on the nonequilbrium Green's function and the generalized Kadanoff--Baym ansatz showed that the excited EI system may undergo a transition towards the normal state when coupled to a bath. However, the isolated EI system perturbed by a laser pulse showed persistent oscillations in the excitonic order parameter but the excitonic order was found to not melt completely. The analysis of the absorbed energy showed a good agreement between the GKBA and KBE in the weak photo-excitation regime. However, for strong excitations the GKBA underestimate the energy absorbed by the pulse.

The character of the dynamics of the EI system, whether excited by a laser pulse or coupled to a bath, was attributed to the narrow spectral features of the GKBA formalism where no proper thermalization channel was found to be present for isolated systems, at least on the level of Hartree--Fock propagators. The bath introduces a suitable decay channel, and we identified separate decay timescales for the excitonic order parameter related to dephasing and thermalization. While we have concentrated on the EI system, we expect our findings to also be general for other symmetry-broken or ordered-phase systems, including e.g., superconducting~\cite{Kemper2015,Sentef2017,Babadi2017,Murakami2017b,Mazza2017b} or charge-density-wave order~\cite{Shen2014,Huber2014,Schueler2018}.

The present implementation of the interacting system embedded in a bath environment, and the subsequent solution of the dynamical equations of motion of the NEGF at the level of the GKBA allows for addressing simultaneously long timescales and large systems. For future work, we therefore highlight the possibility of investigating time-resolved quantum transport in relatively large junctions with electronic correlations~\cite{Latini2014,Bostrom2018,Hopjan2018}. In addition, addressing these effects could provide another route for strong indications of exciton condensation since enhanced tunneling currents in electron-hole double bilayer sheets of graphene and transition-metal dichalcogenide have recently been observed~\cite{Burg2018,Wang2019,Efimkin2020}. The GKBA approach for time-resolved quantum transport could also prove pivotal in, e.g., addressing transiently emerging topological phenomena in Majorana tunnel junctions~\cite{Tuovinen2019njp} with long-lasting characteristic current oscillations. 

\acknowledgments
This research was funded by the Academy of Finland Project No. 321540 (R.T.), 
and the DFG Grant No. SE 2558/2-1 through the Emmy Noether program (M.A.S.). 
We also wish to acknowledge CSC -- IT Center for Science, Finland, for computational resources. The Flatiron Institute is a division of the Simons Foundation.

%\bibliography{refs}


%%%%%%%%%%%%%%%%%%%%%%%%%%%%%%%%%%%%%%%%%%%%%%%%%%%%%%%%%%%%%%%%%%%%%%%%%%%%%%%%
%%%%%%%%%%%%%%%%%%%%%%%%%%%%%%%%%%%%%%%%%%%%%%%%%%%%%%%%%%%%%%%%%%%%%%%%%%%%%%%%

% End document
\end{document}